\documentclass[twocolumn,aps,pra,amssymb,amstext,showpacs,longbibliography]{revtex4-1}
\usepackage{epsf}
\usepackage{subfigure}
\usepackage[colorlinks, linkcolor =blue]{hyperref} 
\usepackage{amsmath}
\usepackage{amssymb}
\usepackage{graphicx}
\usepackage{epstopdf}
\usepackage{color}

\newcommand{\ket}[1]{\left\lvert #1 \right\rangle}
\newcommand{\bol}[1]{\boldsymbol{#1}}

\begin{document}


\title{Quantum backflow for many-particle systems}

\author{M. Barbier}
\affiliation{Center for Nonlinear Phenomena and Complex Systems,\\
Universit\'e Libre de Bruxelles (ULB), Code Postal 231, Campus Plaine, 
B-1050 Brussels, Belgium}


\begin{abstract}
Quantum backflow is the classically-forbidden effect pertaining to the fact that a particle with a positive momentum may exhibit a negative probability current at some space-time point. We investigate how this peculiar phenomenon extends to many-particle systems. We give a general formulation of quantum backflow for systems formed of $N$ free nonrelativistic structureless particles, either identical or distinguishable. Restricting our attention to bosonic systems where the $N$ identical bosons are in the same one-particle state allows us in particular to analytically show that the maximum achievable amount of quantum backflow in this case becomes arbitrarily small for large values of $N$.
\end{abstract}

\noindent 
\vskip 0.5 cm

\maketitle


\section{Introduction}
\label{intro_sec}

The quantum nature of matter challenges our classical intuition through counter-intuitive effects such as diffraction, tunneling or entanglement. An other classically-forbidden phenomenon is quantum backflow \cite{BM94,MB98_Dirac,MB98,EFV05,PGK06,Ber10,Str12,YHH12,PTM13,HGL13,AGP16,BCL17,Gou19,ALS19,DT19,Gou20,MM20,MM20_2part}. The latter stems from the possibility, for a quantum particle following a one-dimensional motion along the $x$-axis, that the probability current at position $x_0$ takes negative values over some time interval even though the particle has a positive momentum. In other words, the probability of finding the particle at positions $-\infty < x < x_0$ may increase over a certain time interval, even though the particle moves in the direction of increasing $x$.

This peculiar effect has first been noted in the context of quantum arrival times \cite{All69}. Its first in-depth study was then performed by Bracken and Melloy \cite{BM94}. In particular, they provided the first evidence of the occurrence of quantum backflow for normalizable wave functions in the case of a free particle. Furthermore, they showed that the magnitude of this effect is limited by a non-trivial upper bound now commonly referred to as the Bracken-Melloy constant. The latter hence quantifies the maximum increase of the probability of finding the particle at positions $-\infty < x < x_0$ that is achievable with positive-momentum states. To date, no analytical expression of this constant has been found but numerical estimations have been obtained \cite{BM94,EFV05,PGK06} with increasing accuracy.

A noteworthy feature of the Bracken-Melloy constant is that it has been shown \cite{BM94} to be a dimensionless quantity that is independent of the duration of the backflow phenomenon, of the mass $m$ of the particle as well as of the (reduced) Planck constant $\hbar$. Therefore, quantum backflow stands as an intrinsically quantum effect that is apparently independent of $\hbar$. This surprising aspect motivated further investigations in order to better understand the fundamental nature of this peculiar phenomenon \cite{PGK06,Ber10,YHH12,HGL13,AGP16}. In particular, the classical limit of quantum backflow remains to be fully comprehended, as the naive classical limit $\hbar \to 0$ clearly can not be readily taken \cite{YHH12}.

While quantum backflow was originally considered in the case of a nonrelativistic free particle, it has ever since been extended to a broad class of other quantum systems. Indeed, it has been shown to occur for a particle in linear \cite{MB98} as well as short-range potentials \cite{BCL17} or for a relativistic free particle \cite{MB98_Dirac,ALS19}. Furthermore, effects akin to quantum backflow have been demonstrated for a nonrelativistic electron in a constant magnetic field \cite{Str12}, the decay of a quasistable system \cite{DT19} or for a dissipative system \cite{MM20}. In addition, a deep connection between quantum backflow and more general classically-forbidden phenomena has been put forward \cite{Gou19,Gou20}. It is also worth stressing that backflow can emerge in other wave phenomena such as in optics \cite{Ber10}. Optical backflow has thus been observed very recently \cite{EZB20}. While a practical scheme based on Bose-Einstein condensates has been proposed \cite{PTM13}, an experimental evidence of backflow on a quantum system still remains to be performed.

Quantum backflow in the context of the time-dependent Schr\"odinger equation has, to the best of our knowledge, been studied exclusively for single-particle systems. Only very recently has backflow been analyzed for a dissipative system of two identical quantum particles coupled to an environment \cite{MM20_2part}.

Therefore, in this work we propose to study the problem of quantum backflow for many-particle systems governed by the time-dependent Schr\"odinger equation. Our aim is thus twofold. On the one hand, we give the first general formulation of quantum backflow for a system formed of $N$ identical particles, either bosons or fermions. This formulation can be easily extended to the case of distinguishable particles. On the other hand, we approach the question of the classical limit of this phenomenon not from the naive limit $\hbar \to 0$ but rather from the limit $N \to \infty$ of a large system. To be more explicit, we show that, in the particular case of $N$ bosons in the same one-particle state, quantum backflow vanishes in the latter limit.

This paper is structured as follows. We begin in section~\ref{single_part_sec} with a brief review of single-particle quantum backflow. This allows us to recall how the latter is quantitatively defined, as well as to fix some notations. We then turn our attention to many-particle quantum backflow in section~\ref{many_part_sec}. Here we give a general formulation of the problem, and illustrate some of the features of backflow in the case of a $N$-boson system. Concluding remarks are finally discussed in section~\ref{A}.


\section{Single-particle quantum backflow}
\label{single_part_sec}

In this section we recall some standard results about the phenomenon of quantum backflow for a single particle. We begin by fixing notations that are used throughout this paper. We consider a nonrelativistic structureless quantum particle of mass $m$ that follows a free one-dimensional motion along the $x$-axis. The dynamical state $\ket{\psi^{(1)} (t)}$ of the system at some time $t \geqslant 0$ is, in the position representation, described by a wave function $\psi^{(1)}(x,t) \equiv \left\langle x | \psi^{(1)}(t) \right\rangle$ that obeys the free-particle time-dependent Schr\"odinger equation
\begin{align}
i \hbar \frac{\partial}{\partial t} \psi^{(1)}(x,t) = - \frac{\hbar^2}{2 m} \frac{\partial^2}{\partial x^2} \psi^{(1)}(x,t) \, .
\label{TDSE_1_part}
\end{align}
This wave function characterizes a probability density $\left\lvert \psi^{(1)}(x,t) \right\rvert^2$ that is required to satisfy the normalization property
\begin{align}
\int_{\mathbb{R}} dx \left\lvert \psi^{(1)}(x,t) \right\rvert^2 = 1 \, .
\label{normalization_1_part}
\end{align}
The latter can be for instance rewritten as
\begin{align}
\int_{-\infty}^{x_0} dx \left\lvert \psi^{(1)}(x,t) \right\rvert^2 + \int_{x_0}^{\infty} dx \left\lvert \psi^{(1)}(x,t) \right\rvert^2 = 1 \, ,
\label{normalization_1_part_alt}
\end{align}
with $x_0$ an arbitrary real number. The first (resp. second) term in the left-hand side of~\eqref{normalization_1_part_alt} merely corresponds to the probability of finding the particle in the position interval $-\infty < x < x_0$ (resp. $x_0 < x < \infty$) at time $t$.

It is worth noting that for a free particle no particular position $x_0$ is privileged. Therefore, without any loss of generality we take for simplicity $x_0=0$ in the sequel. Introducing the notation
\begin{align}
\mathbb{R}^{\pm} \equiv \left\{ x \in \mathbb{R} \quad \vert \quad \mathrm{sgn}(x) = \pm 1 \right\} \, ,
\label{R_pm_def}
\end{align}
with $\mathrm{sgn}(x) \equiv x/\lvert x \rvert$ the sign function, we hence define the probabilities
\begin{align}
\mathcal{P}_{1}^{(1)}(t) \equiv \int_{\mathbb{R}^-} dx \left\lvert \psi^{(1)}(x,t) \right\rvert^2
\label{P_1_1_def}
\end{align}
and
\begin{align}
\mathcal{P}_{0}^{(1)}(t) \equiv \int_{\mathbb{R}^+} dx \left\lvert \psi^{(1)}(x,t) \right\rvert^2
\label{P_0_1_def}
\end{align}
of finding the particle at negative and positive, respectively, positions at time $t$. By construction, the latter correspond to mutually exclusive events and satisfy the normalization condition
\begin{align}
\mathcal{P}_{1}^{(1)}(t) + \mathcal{P}_{0}^{(1)}(t) = 1 \, ,
\label{normalization_1_part_probas}
\end{align}
as a direct consequence of~\eqref{normalization_1_part_alt} for $x_0 = 0$.

In addition to the probability density $\left\lvert \psi^{(1)}(x,t) \right\rvert^2$, one can also consider the probability current $\mathcal{J}^{(1)}(x,t)$ defined by
\begin{multline}
\mathcal{J}^{(1)}(x,t) \equiv -i \frac{\hbar}{2 m} \left[ \psi^{(1)^*}(x,t) \frac{\partial}{\partial x} \psi^{(1)}(x,t) \right. \\
- \left. \psi^{(1)}(x,t) \frac{\partial}{\partial x} \psi^{(1)^*}(x,t) \right] \, ,
\label{current_1_part}
\end{multline}
where $z^*$ denotes the complex conjugate of the complex number $z$. The probability density and current satisfy the conservation equation
\begin{align}
\frac{\partial}{\partial t} \left\lvert \psi^{(1)}(x,t) \right\rvert^2 + \frac{\partial}{\partial x} \mathcal{J}^{(1)}(x,t) = 0
\label{cons_eq_1_part}
\end{align}
as a direct consequence of the Schr\"odinger equation~\eqref{TDSE_1_part}. Differentiating~\eqref{P_1_1_def} with respect to time and using~\eqref{cons_eq_1_part} hence shows, also using~\eqref{normalization_1_part_probas}, that
\begin{align}
\frac{d \mathcal{P}_{1}^{(1)}}{dt} = - \mathcal{J}^{(1)}(0,t) = - \frac{d \mathcal{P}_{0}^{(1)}}{dt} \, ,
\label{deriv_P_1_1_j_0}
\end{align}
where we used the fact that $\lim_{x \to \pm \infty} \mathcal{J}^{(1)}(x,t)=0$. Indeed, the wave function itself must vanish at infinity, which ensures that the probability of finding the particle at infinity vanishes at any finite time $t$. It is worth stressing that~\eqref{deriv_P_1_1_j_0} is peculiar to the one-dimensional motion of a single particle, as the conservation equation takes the particularly simple form~\eqref{cons_eq_1_part} in this case. Such a simple relation between time-derivatives of the probabilities and the current can not be established for a many-particle system, as is seen in more details in section~\ref{many_part_sec} below.

After this reminder of the quantum mechanical description of a free particle, we now give in subsection~\ref{backflow_subsec} a short outline of single-particle quantum backflow. An explicit example that is known to give rise to backflow is then reviewed in subsection~\ref{example_subsec} to illustrate this peculiar quantum effect.


\subsection{Quantum backflow}
\label{backflow_subsec}

The phenomenon of quantum backflow is rooted in the existence of states $\psi^{(1)}(x,t)$ that make the probability $\mathcal{P}_{1}^{(1)}(t)$ increase over some time interval \textit{even though} the particle has a positive momentum. Such a behavior is clearly impossible from the classical-mechanical point of view. Indeed, if a classical free particle has a positive (though uncertain) velocity, the probability of finding it on the negative $x$-axis can be shown to be a monotonically decreasing function of time \cite{BM94}.

The idea of a quantum particle with a positive momentum can be made precise by writing the Fourier transform of the wave function $\psi^{(1)}(x,t)$, which is thus required to contain only positive components of the momentum $p$. That is, the particle is assumed to be prepared in the initial state
\begin{align}
\psi^{(1)}(x,0) = \frac{1}{\sqrt{2 \pi \hbar}} \int_{\mathbb{R}^+} dp \, \mathrm{e}^{ixp/\hbar} \phi^{(1)}(p) \, ,
\label{psi_0_def}
\end{align}
where the functions $\psi^{(1)}(x,0)$ and $\phi^{(1)}(p)$ both satisfy the normalization condition
\begin{align}
\int_{\mathbb{R}} dx \left\lvert \psi^{(1)}(x,0) \right\rvert^2 = \int_{\mathbb{R}^+} dp \left\lvert \phi^{(1)}(p) \right\rvert^2 = 1 \, .
\label{norm_cond_psi_phi}
\end{align}
It is worth noting that the restriction of the momentum integral to $\mathbb{R}^+$ is ensured by the fact that $\phi^{(1)}(p)=0$ for any $p<0$. This stems from the particular initial state~\eqref{psi_0_def} considered here. Indeed, the normalized wave function $\psi^{(1)}(x,0)$ must admit a decomposition on the basis formed by the eigenvectors of the free-particle Hamiltonian, i.e. precisely the plane waves with both positive \textit{and} negative momenta. To consider a linear superposition~\eqref{psi_0_def} of plane waves with only positive momenta implies that the coefficients of the plane waves with negative momenta all vanish, i.e. $\phi^{(1)}(p)=0$ for any $p<0$.

Now, an important consequence of considering a free particle is that the wave function $\psi^{(1)}(x,t)$ that evolves from the initial state~\eqref{psi_0_def} is of the form
\begin{align}
\psi^{(1)}(x,t) = \frac{1}{\sqrt{2 \pi \hbar}} \int_{\mathbb{R}^+} dp \, \mathrm{e}^{-i p^2 t/2 m \hbar} \, \mathrm{e}^{ixp/\hbar} \phi^{(1)}(p) \, ,
\label{psi_1_t_gen_expr}
\end{align}
as can be easily shown from the Schr\"odinger equation~\eqref{TDSE_1_part}. We emphasize that the integration range in~\eqref{psi_1_t_gen_expr} is again $\mathbb{R}^+$, as in the initial state~\eqref{psi_0_def}. This can be understood from the absence, for a free particle, of a potential that can induce negative momenta, e.g. through the reflection on a barrier. Therefore, the expression~\eqref{psi_1_t_gen_expr} of the wave function $\psi^{(1)}(x,t)$ is the quantum translation of the particle having, with probability one, a positive momentum at any time $t \geqslant 0$.

We now consider the probability $\mathcal{P}_{1}^{(1)}(t)$, as defined by~\eqref{P_1_1_def}, of finding the particle on the negative $x$-axis at time $t$ for a state of the form~\eqref{psi_1_t_gen_expr}. We introduce the change $\Delta_1$ of $\mathcal{P}_{1}^{(1)}$ over a fixed (though arbitrary) time interval $0 \leqslant t \leqslant T$ for some $T>0$, which is defined by
\begin{align}
\Delta_1 \equiv \mathcal{P}_{1}^{(1)}(T) - \mathcal{P}_{1}^{(1)} (0) \, .
\label{Delta_1_def}
\end{align}
This quantity allows to quantitatively study the phenomenon of quantum backflow, which then rises from positive values of $\Delta_1$ \cite{BM94}. Indeed, to have $\Delta_1 > 0$ means that the probability $\mathcal{P}_{1}^{(1)}$ has increased between the times $t=0$ and $t=T$. Note that $\Delta_1$ can, in view of the normalization condition~\eqref{normalization_1_part_probas}, be alternatively written as
\begin{align}
\Delta_1 = \mathcal{P}_{0}^{(1)}(0) - \mathcal{P}_{0}^{(1)} (T) \, .
\label{Delta_1_P_0_def}
\end{align}
Quantum backflow can thus be equivalently viewed as rising from the decrease of the probability $\mathcal{P}_{0}^{(1)}$ of finding the particle on the positive real axis between the times $t=0$ and $t=T$.

It is worth noting that~\eqref{Delta_1_def} can be written in the form
\begin{align}
\Delta_1 = \int_{0}^{T} dt \, \frac{d \mathcal{P}_{1}^{(1)}}{dt} \, .
\label{Delta_1_deriv_P_1_1_rel}
\end{align}
Substituting~\eqref{deriv_P_1_1_j_0} into~\eqref{Delta_1_deriv_P_1_1_rel} hence allows to express $\Delta_1$ in terms of the probability current $\mathcal{J}^{(1)}$ and
\begin{align}
\Delta_1 = - \int_{0}^{T} dt \, \mathcal{J}^{(1)}(0,t) \, .
\label{Delta_1_j_0_expr}
\end{align}
This shows that quantum backflow, i.e. to have $\Delta_1 > 0$, can only occur if the current $\mathcal{J}^{(1)}(0,t)$ takes negative values at some times $0 \leqslant t \leqslant T$. Here again, we emphasize that the relation~\eqref{Delta_1_j_0_expr} is peculiar to the one-dimensional motion of a single particle. In the many-particle case, one must rather extend the original definition~\eqref{Delta_1_def} of $\Delta_1$, as is discussed in section~\ref{many_part_sec} below.

It is clear from the definition~\eqref{Delta_1_def} of $\Delta_1$ as the difference of two probabilities that the latter takes values between $-1$ and 1. Interestingly, it has been found \cite{BM94} that $\Delta_1$ actually admits an upper bound $\Delta_{1,\text{max}}$ that is much stricter than 1. While no exact expression of $\Delta_{1,\text{max}}$ has been obtained to date, numerical investigations have led to the estimate \cite{BM94,EFV05,PGK06}
\begin{align}
\Delta_{1,\text{max}} \approx 0.0384517 \, .
\label{BM_constant}
\end{align}
This is the so-called Bracken-Melloy constant. It quantifies the maximum amount of quantum backflow, that is the maximum increase of the probability of finding the particle on the negative real axis over an arbitrary time interval $0 \leqslant t \leqslant T$ for a positive-momentum state of the form~\eqref{psi_1_t_gen_expr}.

As we indicated above in the introduction, a surprising feature of the Bracken-Melloy constant $\Delta_{1,\text{max}}$ is that it proves to be independent of the time parameter $T$, as well as of the mass $m$ and of the (reduced) Planck constant $\hbar$. This rises from the combined facts that no dimensionless quantity can be constructed from $T$, $m$ and $\hbar$, and that no natural length scale is associated to a free particle. This led to the interpretation of the maximum backflow~\eqref{BM_constant} as a purely quantum effect that is independent of Planck's constant \cite{BM94}.

This observation hence raises the question of the classical limit of quantum backflow, as the naive classical limit $\hbar \to 0$ can not be readily taken. As is discussed in \cite{YHH12}, a possible approach is to consider realistic measurements of the position of the particle at times $t=0$ and $t=T$ modeled by quasiprojectors rather than by projectors. This allows to introduce a length scale in the problem, which represents the precision of the position measurement. The resulting maximum backflow then depends on $\hbar$, and can thus be studied in the naive classical limit $\hbar \to 0$ where it is seen to vanish. As we discuss in section~\ref{many_part_sec} below, to consider a $N$-particle system allows us to approach the question of the classical limit of quantum backflow from a different point of view. In this case, the classical limit can be viewed as the limit of a very large number of particles, i.e. $N \to \infty$.

Various analytical examples of wave functions of the form~\eqref{psi_1_t_gen_expr} that give rise to quantum backflow have been studied \cite{BM94,YHH12,HGL13}. We now recall one such wave function, which we use again in section~\ref{many_part_sec} below to illustrate some features of the phenomenon of quantum backflow in the case of a many-particle system.


\subsection{An explicit backflow state}
\label{example_subsec}

In this section we consider a particular example of wave function of the form~\eqref{psi_1_t_gen_expr} that has been previously discussed in \cite{BM94} in order to explicitly demonstrate the occurrence of quantum backflow for a single free particle.

This example stems from choosing a particular initial momentum wave function $\phi^{(1)}(p)$ in~\eqref{psi_1_t_gen_expr}, namely $\phi^{(1)}(p) = \widetilde{\phi}(p)$ with $\widetilde{\phi}(p)$ given by the superposition of exponentials
\begin{equation}
\widetilde{\phi}(p) \equiv \left\{\begin{array}{ll}
0 \quad & , \quad \text{if} \quad p < 0 \\[0.3cm]
\frac{18}{\sqrt{35 \alpha^3}} \, p \left( \mathrm{e}^{-p/\alpha} - \frac{1}{6} \, \mathrm{e}^{-p/2\alpha} \right) \quad & , \quad \text{if} \quad p > 0
\end{array}\right. \, ,
\label{phi_tilde_def}
\end{equation}
where $\alpha$ is a positive constant that has the dimension of a momentum. Note that the function $\widetilde{\phi}(p)$ is continuous at $p=0$. Substituting~\eqref{phi_tilde_def} into~\eqref{psi_1_t_gen_expr} then expresses the resulting wave function
\begin{widetext}
\begin{align}
\psi^{(1)}(x,t) = \widetilde{\psi} \left( \frac{\alpha x}{\hbar} , \frac{\alpha^2 t}{m \hbar} \right) \equiv \frac{1}{\sqrt{2 \pi \hbar}} \int_{\mathbb{R}^+} dp \, \mathrm{e}^{-i p^2 t/2 m \hbar} \, \mathrm{e}^{ixp/\hbar} \, \widetilde{\phi}(p)
\label{psi_tilde_def}
\end{align}
as a Gaussian integral that can be computed analytically (see e.g. \cite{AbrSteg,GradRyz}), eventually yielding
\begin{multline}
\widetilde{\psi}(x',t') = -18 \sqrt{\frac{\alpha}{70 \pi \hbar}} \left( \frac{5i}{6 t'} + \sqrt{\frac{\pi}{4t'^3}} (i-1) \left\{ (x'+i) \, \mathrm{exp} \left[ \frac{i}{2t'} (x'+i)^2 \right] \mathrm{erfc} \left[ - \frac{(1+i)(x'+i)}{\sqrt{4 t'}} \right] \right. \right. \\[0.5cm]
\left. \left. - \frac{2x'+i}{12} \, \mathrm{exp} \left[ \frac{i}{8 t'} (2x'+i)^2 \right] \mathrm{erfc} \left[ - \frac{(1+i)(2x'+i)}{\sqrt{16t'}} \right] \right\} \right) \, ,
\label{psi_t_error_fct_expr}
\end{multline}
\end{widetext}
with
\begin{align}
\mathrm{erfc}(z) = 1 - \mathrm{erf}(z) = \frac{2}{\sqrt{\pi}} \int_{z}^{\infty} dy \, \mathrm{e}^{-y^2}
\label{erfc_def}
\end{align}
the complementary error function, and where the dimensionless quantities $x'$ and $t'$ are related to the position $x$ and the time $t$ through
\begin{align}
x' \equiv \frac{\alpha x}{\hbar} \qquad \text{and} \qquad t' \equiv \frac{\alpha^2 t}{m \hbar} \, .
\label{dimensionless_x_and_t_def}
\end{align}
We emphasize that, while the wave function $\widetilde{\psi}$ depends on the dimensionless variables $x'$ and $t'$, it has the same dimension as $\psi^{(1)}$ (namely the inverse square root of a length).

The behavior of the wave function $\widetilde{\psi}(x',t')$ close to $t'=0$ can be obtained from~\eqref{psi_t_error_fct_expr} by noting that as $t' \to 0$ the modulus of the arguments of the complementary error functions diverge. For $t' \ll 1$ we can thus substitute the well-known asymptotic expansion (see e.g. \cite{GradRyz})
\begin{align}
\mathrm{erfc}(z) \sim \frac{\mathrm{e}^{-z^2}}{\sqrt{\pi}z} \sum_{k} (-1)^k \frac{(2k-1)!!}{\left( 2z^2 \right)^k} \, ,
\label{erfc_asympt}
\end{align}
with $(2k-1)!! \equiv (2k-1)(2k-3) \cdots \times 3$ the double factorial, into~\eqref{psi_t_error_fct_expr} to get
\begin{widetext}
\begin{align}
\widetilde{\psi}(x',t') \sim 18 \sqrt{\frac{\alpha}{70 \pi \hbar}} \left\{ \frac{1}{(1-ix')^2} - \frac{2}{3} \frac{1}{(1-2ix')^2} + \sum_{k \geqslant 2} (-i t')^{k-1} (2k-1)!! \left[ \frac{1}{(1-ix')^{2k}} - \frac{1}{6} \left( \frac{1}{1-2ix'} \right)^{2k} \right] \right\}
\label{psi_t_asympt_expr}
\end{align}
\end{widetext}
in the vicinity of $t'=0$. Setting $t'=0$ into~\eqref{psi_t_asympt_expr} readily yields the initial state $\widetilde{\psi}(x',0)$.

One can then compute the corresponding probability current $\mathcal{J}^{(1)}(x,t) = \widetilde{\mathcal{J}}(x',t')$ , which in view of the definition~\eqref{current_1_part} is given by
\begin{align}
\widetilde{\mathcal{J}}(x',t') = -i \frac{\alpha}{2 m} \left[ \widetilde{\psi}^{*} \frac{\partial \widetilde{\psi}}{\partial x'} - \widetilde{\psi} \, \frac{\partial \widetilde{\psi}^{*}}{\partial x'} \right] \, ,
\label{J_tilde_def}
\end{align}
where we used~\eqref{dimensionless_x_and_t_def}. Substituting the expression~\eqref{psi_t_asympt_expr} of $\widetilde{\psi}$ for $t'=0$ into~\eqref{J_tilde_def} and setting $x'=0$ into the resulting expression of $\widetilde{\mathcal{J}}$ hence yields \cite{BM94}
\begin{align}
\widetilde{\mathcal{J}}(0,0) = - \frac{36 \alpha^2}{35 \pi m \hbar} \, ,
\label{initial_neg_current}
\end{align}
which is clearly negative. 

Furthermore, the probabilities $\mathcal{P}_{0,1}^{(1)}(t)=\widetilde{\mathcal{P}}_{0,1}^{(1)}(t')$ are here obtained by merely substituting $\psi^{(1)} = \widetilde{\psi}$ into the definitions~\eqref{P_1_1_def}-\eqref{P_0_1_def} and we have, again using~\eqref{dimensionless_x_and_t_def},
\begin{align}
\widetilde{\mathcal{P}}_{1}^{(1)}(t') = \frac{\hbar}{\alpha} \int_{\mathbb{R}^-} dx' \left\lvert \widetilde{\psi}(x',t') \right\rvert^2
\label{P_1_1_example}
\end{align}
and
\begin{align}
\widetilde{\mathcal{P}}_{0}^{(1)}(t') = \frac{\hbar}{\alpha} \int_{\mathbb{R}^+} dx' \left\lvert \widetilde{\psi}(x',t') \right\rvert^2 \, .
\label{P_0_1_example}
\end{align}
Combining~\eqref{deriv_P_1_1_j_0} with~\eqref{dimensionless_x_and_t_def} and~\eqref{initial_neg_current} then readily yields \cite{BM94}
\begin{align}
\left. \frac{d \widetilde{\mathcal{P}}_{1}^{(1)}}{dt'} \right\rvert_{t'=0} = - \left. \frac{d \widetilde{\mathcal{P}}_{0}^{(1)}}{dt'} \right\rvert_{t'=0} = \frac{36}{35 \pi} > 0 \, .
\label{deriv_P_1_1_t_0}
\end{align}
Therefore, the probability of finding the particle described by the wave function~\eqref{psi_tilde_def} on the negative real axis initially increases, even though the particle has a positive momentum. This indeed demonstrates the occurrence of the phenomenon of quantum backflow.

A numerical analysis shows \cite{BM94} that the derivative $d \widetilde{\mathcal{P}}_{1}^{(1)}/dt'$ remains positive for times $t'$ ranging between 0 and $t' = t'_1 \approx 0.021$. This means that the probability of finding the particle on the negative real axis increases by a maximum amount $\widetilde{\Delta}_{1,\text{max}}$ here given by
\begin{align}
\widetilde{\Delta}_{1,\text{max}} = \widetilde{\mathcal{P}}_{1}^{(1)}(t'_1) - \widetilde{\mathcal{P}}_{1}^{(1)} (0) \, ,
\label{Delta_def}
\end{align}
which can be numerically evaluated to
\begin{align}
\widetilde{\Delta}_{1,\text{max}} \approx 0.0043 \, ,
\label{Delta_expr}
\end{align}
that is approximately 11\% of the maximum achievable backflow quantified by the Bracken-Melloy constant~\eqref{BM_constant}.

We recalled in this section some standard results about quantum backflow for a single particle. In particular, we saw that it can be adequately quantified by the probability change $\Delta_1$ defined by~\eqref{Delta_1_def}. Since the latter admits the upper bound~\eqref{BM_constant}, there is a fundamental limit to the maximum amount of backflow for a single particle. We now discuss how quantum backflow extends to many-particle systems.


\section{Many-particle quantum backflow}
\label{many_part_sec}

In this section we study the phenomenon of quantum backflow in the case of a many-particle system. Our main aim is to investigate the behavior of the former with respect to the number $N$ of particles.

To this end, we propose in subsection~\ref{gen_formulation_subsec} a general formulation of the problem. We then restrict our attention to the particular case of a system formed of $N$ bosons that are all in a same one-particle state. As is seen in subsection~\ref{N_bos_subsec}, this assumption allows us to express the quantities of interest in terms of the underlying single-particle ones. We can thus build up on the physical intuition gained from the single-particle case, and we show in subsection~\ref{class_limit_subsec} that quantum backflow vanishes for a large number $N$ of bosons. These conclusions are then illustrated in subsection~\ref{example_N_bos_subsec} by means of the explicit example that we discussed in the previous section.


\subsection{General formulation}
\label{gen_formulation_subsec}

We consider a system of $N$ identical nonrelativistic structureless quantum particles of mass $m$, with $N \geqslant 1$. The particles are assumed to propagate freely at one dimension. For compactness we introduce the $N$-component vector $\bol{x}$ defined by
\begin{align}
\bol{x} \equiv \left( x_1, \ldots , x_N \right) \, ,
\label{x_vector_def}
\end{align}
hence representing the position vector of the $N$-particle system. Differential elements in $N$-dimensional integrals are then merely denoted by $d\bol{x} \equiv dx_1 \cdots dx_N$.

The dynamical state $\ket{\psi^{(N)} (t)}$ of the system at some time $t \geqslant 0$ is thus, in the position representation, described by a wave function $\psi^{(N)}(\bol{x},t) \equiv \left\langle \bol{x} | \psi^{(N)}(t) \right\rangle$ that obeys the free $N$-particle time-dependent Schr\"odinger equation
\begin{align}
i \hbar \frac{\partial}{\partial t} \psi^{(N)}(\bol{x},t) = - \frac{\hbar^2}{2 m} \sum_{j=1}^{N} \frac{\partial^2}{\partial x_j^2} \, \psi^{(N)}(\bol{x},t) \, .
\label{TDSE_N_part}
\end{align}
The resulting probability density $\left\lvert \psi^{(N)}(\bol{x},t) \right\rvert^2$ is assumed to be normalized, i.e.
\begin{align}
\int_{\mathbb{R}^N} d\bol{x} \left\lvert \psi^{(N)}(\bol{x},t) \right\rvert^2 = 1 \, .
\label{normalization_N_part}
\end{align}
The wave function $\psi^{(N)}$ also characterizes a probability current $\bol{\mathcal{J}}^{(N)}(\bol{x},t)$, which is now a vector quantity, defined by
\begin{multline}
\bol{\mathcal{J}}^{(N)}(\bol{x},t) \equiv -i \frac{\hbar}{2 m} \left[ \psi^{(N)^*}(\bol{x},t) \bol{\nabla} \psi^{(N)}(\bol{x},t) \right. \\
- \left. \psi^{(N)}(\bol{x},t) \bol{\nabla} \psi^{(N)^*}(\bol{x},t) \right]
\label{current_N_part}
\end{multline}
in terms of the gradient operator
\begin{align}
\bol{\nabla} \equiv \sum_{j=1}^{N} \hat{x}_j \frac{\partial}{\partial x_j} \, ,
\label{nabla_def}
\end{align}
where the vectors $\hat{x}_1, \ldots , \hat{x}_N$ form an orthonormal basis of $\mathbb{R}^N$, i.e. $\hat{x}_j \cdot \hat{x}_k = \delta_{jk}$ with $\delta_{jk}$ the Kronecker delta. The probability density and the current still satisfy a conservation equation, here given by
\begin{align}
\frac{\partial}{\partial t} \left\lvert \psi^{(N)}(\bol{x},t) \right\rvert^2 + \bol{\nabla} \cdot \bol{\mathcal{J}}^{(N)}(\bol{x},t) = 0 \, ,
\label{cons_eq_N_part}
\end{align}
as a direct consequence of the Schr\"odinger equation~\eqref{TDSE_N_part}.

Similarly to the single-particle case, we again assume that the particles are initially prepared with positive momenta. The $N$-particle wave function $\psi^{(N)}$ can thus be written in the form~\eqref{psi_1_t_gen_expr}, that is here
\begin{multline}
\psi^{(N)}(\bol{x},t) = \frac{1}{\left( 2 \pi \hbar \right)^{N/2}} \int_{\left( \mathbb{R}^+ \right)^N} d\bol{p} \, \mathrm{e}^{i\bol{x} \cdot \bol{p}/\hbar} \\
\times \mathrm{e}^{-i \bol{p}^2 t/2 m \hbar} \, \phi^{(N)}(\bol{p}) \, ,
\label{psi_N_t_gen_expr}
\end{multline}
with the $N$-component vector
\begin{align}
\bol{p} \equiv \left( p_1, \ldots , p_N \right)
\label{p_vector_def}
\end{align}
representing the momentum of the $N$-particle system, and where the differential element is merely $d\bol{p} \equiv dp_1 \cdots dp_N$. We readily recover the single-particle wave function~\eqref{psi_1_t_gen_expr} upon setting $N=1$ into~\eqref{psi_N_t_gen_expr}. The $N$-particle momentum wave function $\phi^{(N)}$ is thus itself normalized in view of~\eqref{normalization_N_part}, i.e.
\begin{align}
\int_{\left( \mathbb{R}^+ \right)^N} d\bol{p} \left\lvert \phi^{(N)}(\bol{p}) \right\rvert^2 = 1 \, .
\label{normalization_phi_N}
\end{align}

Now, inspired by the single-particle probabilities $\mathcal{P}_{1}^{(1)}$ and $\mathcal{P}_{0}^{(1)}$ defined by~\eqref{P_1_1_def}-\eqref{P_0_1_def}, we introduce the probabilities $\mathcal{P}_{j}^{(N)}(t)$, $j=0, \ldots , N$, of finding $j$ of the $N$ particles on the negative real axis, and thus the remaining $N-j$ particles on the positive real axis, at time $t$. Since these probabilities refer to mutually exclusive events, we must have the normalization condition
\begin{align}
\sum_{j=0}^{N} \mathcal{P}_{j}^{(N)}(t) = 1
\label{normalization_N_part_probas}
\end{align}
for any $t \geqslant 0$. The expression of these probabilities can e.g. be obtained from~\eqref{normalization_N_part} by successively splitting each integral over $\mathbb{R}$ as one integral over $\mathbb{R}^-$ and one over $\mathbb{R}^+$, hence yielding
\begin{widetext}
\begin{align*}
1 & = \int_{\mathbb{R}^{N-1}} dx_1 \cdots dx_{N-1} \int_{\mathbb{R}^{-}} dx_N \left\lvert \psi^{(N)} \right\rvert^2 + \int_{\mathbb{R}^{N-1}} dx_1 \cdots dx_{N-1} \int_{\mathbb{R}^{+}} dx_N \left\lvert \psi^{(N)} \right\rvert^2 \\
& = \int_{\mathbb{R}^{N-2}} dx_1 \cdots dx_{N-2} \int_{\mathbb{R}^{-}} dx_{N-1} \int_{\mathbb{R}^{-}} dx_N \left\lvert \psi^{(N)} \right\rvert^2 + \int_{\mathbb{R}^{N-2}} dx_1 \cdots dx_{N-2} \int_{\mathbb{R}^{+}} dx_{N-1} \int_{\mathbb{R}^{-}} dx_N \left\lvert \psi^{(N)} \right\rvert^2 \\
& + \int_{\mathbb{R}^{N-2}} dx_1 \cdots dx_{N-2} \int_{\mathbb{R}^{-}} dx_{N-1} \int_{\mathbb{R}^{+}} dx_N \left\lvert \psi^{(N)} \right\rvert^2 + \int_{\mathbb{R}^{N-2}} dx_1 \cdots dx_{N-2} \int_{\mathbb{R}^{+}} dx_{N-1} \int_{\mathbb{R}^{+}} dx_N \left\lvert \psi^{(N)} \right\rvert^2 \\
& = \cdots = \sum_{j=0}^{N} \mathcal{P}_{j}^{(N)}(t) \, ,
\end{align*}
where $\mathcal{P}_{j}^{(N)}$ is thus given by
\begin{multline}
\mathcal{P}_{j}^{(N)}(t) = \sum_{k_1=1}^{N} \sum_{\substack{k_2=1 \\ k_2>k_1}}^{N} \cdots \sum_{\substack{k_j=1 \\ k_j>k_{j-1}}}^{N} \int_{\mathbb{R}^{+}} dx_{1} \cdots \int_{\mathbb{R}^{+}} dx_{k_1-1} \int_{\mathbb{R}^{-}} dx_{k_1} \int_{\mathbb{R}^{+}} dx_{k_1+1} \cdots \\
\times \int_{\mathbb{R}^{+}} dx_{k_j-1} \int_{\mathbb{R}^{-}} dx_{k_j} \int_{\mathbb{R}^{+}} dx_{k_j+1} \cdots \int_{\mathbb{R}^{+}} dx_{N} \left\lvert \psi^{(N)}(\bol{x},t) \right\rvert^2 \, .
\label{P_j_N_def}
\end{multline}
\end{widetext}
This definition remains valid for $j=0$ if we agree that in this case the integration domain is merely $\left( \mathbb{R}^+ \right)^N$.

In addition to $\mathcal{P}_{j}^{(N)}$, we also define the probability $\mathcal{P}_{-}^{(N)}(t)$ by
\begin{align}
\mathcal{P}_{-}^{(N)}(t) \equiv \sum_{j=1}^{N} \mathcal{P}_{j}^{(N)}(t) \, .
\label{P_min_def}
\end{align}
Since the probabilities $\mathcal{P}_{j}^{(N)}$ refer to mutually exclusive events, $\mathcal{P}_{-}^{(N)}(t)$ hence corresponds to the probability of finding \textit{at least} one particle on the negative real axis at time $t$. In the single-particle case, it merely corresponds to the probability $\mathcal{P}_{1}^{(1)}$, i.e. $\mathcal{P}_{-}^{(1)} = \mathcal{P}_{1}^{(1)}$, as can be readily seen upon setting $N=1$ into~\eqref{P_min_def}. It is also worth noting that combining~\eqref{P_min_def} with the normalization condition~\eqref{normalization_N_part_probas} immediately shows that $\mathcal{P}_{-}^{(N)}$ can be alternatively written as
\begin{align}
\mathcal{P}_{-}^{(N)}(t) = 1 - \mathcal{P}_{0}^{(N)}(t) \, .
\label{P_BF_P_0_N_rel}
\end{align}
This form highlights the fact that $\mathcal{P}_{-}^{(N)}$ and $\mathcal{P}_{0}^{(N)}$ refer to complementary events, namely to find or not to find, respectively, a particle on the negative real axis.

Finally, we introduce the quantity $\Delta_N$ that generalizes its single-particle counterpart $\Delta_1$ defined by~\eqref{Delta_1_def}. We recall that the latter characterizes the change of the probability $\mathcal{P}_{1}^{(1)}$, i.e. $\mathcal{P}_{-}^{(1)}$, of finding the particle on the negative real axis over a fixed but arbitrary time interval $0 \leqslant t \leqslant T$, for some $T>0$. Therefore, we propose to define the quantity $\Delta_N$ by
\begin{align}
\Delta_N \equiv \mathcal{P}_{-}^{(N)}(T) - \mathcal{P}_{-}^{(N)}(0) \, ,
\label{Delta_N_def}
\end{align}
which immediately gives back the definition~\eqref{Delta_1_def} of $\Delta_1$ for $N=1$. Note that~\eqref{Delta_N_def} can also be equivalently written as
\begin{align}
\Delta_N = \mathcal{P}_{0}^{(N)}(0) - \mathcal{P}_{0}^{(N)}(T)
\label{Delta_N_def_P_0}
\end{align}
in view of~\eqref{P_BF_P_0_N_rel}.

We believe that the quantity $\Delta_N$ defined by~\eqref{Delta_N_def} or, equivalently, by~\eqref{Delta_N_def_P_0} is the natural quantifier of quantum backflow for a $N$-particle system. Indeed, remember that for a single particle with a positive momentum backflow rises from the non-classical fact that the probability of finding the particle on the positive real axis may decrease over the time interval $0 \leqslant t \leqslant T$. The equivalent for a system of $N$ particles with positive momenta must thus be that the probability $\mathcal{P}_{0}^{(N)}$ of finding \textit{all} particles on the positive real axis possibly decreases between the times $t=0$ and $t=T$. Such a decrease of $\mathcal{P}_{0}^{(N)}$ can be viewed as resulting from having at least one of the $N$ particles traveling backwards from an initial positive position to a negative one, which precisely corresponds to the physical intuition that underlies the idea of backflow. Similarly to the single-particle case, the occurrence of quantum backflow for a $N$-particle system is thus embedded into the positive values of the quantity $\Delta_N$.

It is here worth emphasizing that the simple relation~\eqref{deriv_P_1_1_j_0} between the time derivatives of $\mathcal{P}_{1}^{(1)}$ and $\mathcal{P}_{0}^{(1)}$ and the current $\mathcal{J}^{(1)}$ can not be extended to a $N$-particle system. Indeed, setting e.g. $j=0$ into~\eqref{P_j_N_def}, differentiating with respect to time and using the conservation equation~\eqref{cons_eq_N_part} yields
\begin{align}
\frac{d \mathcal{P}_{0}^{(N)}}{dt} = - \int_{\left( \mathbb{R}^+ \right)^N} d\bol{x} \, \bol{\nabla} \cdot \bol{\mathcal{J}}^{(N)}(\bol{x},t) \, .
\label{deriv_P_0_N}
\end{align}
While $\nabla \mathcal{J}^{(1)}(x,t)$ can be easily integrated with respect to the position $x$, this is not the case of $\bol{\nabla} \cdot \bol{\mathcal{J}}^{(N)}(\bol{x},t)$ for $N>1$. By extension, this also precludes a simple relation of the form~\eqref{Delta_1_j_0_expr} [which was a direct consequence of~\eqref{deriv_P_1_1_j_0}] between the probability change $\Delta_N$ and the current $\bol{\mathcal{J}}^{(N)}$.

The above formulation applies to a general system formed of $N$ identical free particles, the latter being either bosons or fermions. It can also be straightforwardly extended to the case of distinguishable particles with different masses $m_j$, $j=1, \ldots, N$. We now discuss how the problem can be simplified in the case of $N$ bosons that are all in the same one-particle state.


\subsection{Bosonic system}
\label{N_bos_subsec}

From now on we assume that the $N$-particle system consists of $N$ identical bosons that are all in the same one-particle state $\psi^{(1)}(x,t)$. Therefore, the $N$-particle wave function $\psi^{(N)}$ can be written as the mere product state
\begin{align}
\psi^{(N)}(\bol{x},t) = \prod_{j=1}^{N} \psi^{(1)}(x_j,t) \, ,
\label{psi_N_prod_state_def}
\end{align}
while the corresponding initial momentum wave function $\phi^{(N)}$ reads
\begin{align}
\phi^{(N)}(\bol{p}) = \prod_{j=1}^{N} \phi^{(1)}(p_j) \, ,
\label{phi_N_prod_state_def}
\end{align}
as can be seen upon substituting the expression~\eqref{psi_1_t_gen_expr} of $\psi^{(1)}$ into~\eqref{psi_N_prod_state_def} and comparing the resulting expression of $\psi^{(N)}$ to its Fourier transform~\eqref{psi_N_t_gen_expr}. The normalization conditions~\eqref{normalization_N_part} and~\eqref{normalization_phi_N} are then direct consequences of their single-particle counterparts~\eqref{normalization_1_part} and~\eqref{norm_cond_psi_phi}, respectively.

To focus on the simple product states~\eqref{psi_N_prod_state_def} certainly reduces the space of $N$-particle states that we consider. Such an assumption is however well justified in view of experiments based on cold atoms or Bose-Einstein condensates (see e.g. \cite{Ing} as a general reference). Indeed, suppose that a Bose-Einstein condensate is prepared at a sufficiently low temperature and that the bosons can be treated as independent, i.e. particle-particle interactions are neglected. Then the state of the condensed bosons can be, to a good approximation (the lower the temperature, the better the approximation), described by a pure state that is precisely of the form~\eqref{psi_N_prod_state_def}. In such a case, the initial one-particle state $\psi^{(1)}(x,0)$ corresponds to the ground state of the single-particle Hamiltonian that is used to trap the bosons.

In addition to being practically relevant, the product state~\eqref{psi_N_prod_state_def} allows to greatly simplify our formulation of many-particle quantum backflow. Indeed, substituting first~\eqref{psi_N_prod_state_def} into the definition~\eqref{P_j_N_def} allows to factorize the $N$-particle probability $\mathcal{P}_{j}^{(N)}$ as
\begin{multline}
\mathcal{P}_{j}^{(N)}(t) = \left[ \int_{\mathbb{R}^{-}} dx \left\lvert \psi^{(1)}(x,t) \right\rvert^2 \right]^{j} \\
\times \left[ \int_{\mathbb{R}^{+}} dx \left\lvert \psi^{(1)}(x,t) \right\rvert^2 \right]^{N-j} \sum_{k_1=1}^{N} \sum_{\substack{k_2=1 \\ k_2>k_1}}^{N} \cdots \sum_{\substack{k_j=1 \\ k_j>k_{j-1}}}^{N} 1 \, .
\label{P_j_N_prod_state}
\end{multline}

The nested sum in the right-hand side of~\eqref{P_j_N_prod_state} can be evaluated as follows. Consider the set $\mathcal{S}^{(N)} \equiv \{ 1, \ldots , N \}$ of $N$ elements. To compute the sum in~\eqref{P_j_N_prod_state} is thus equivalent to determining the total number of subsets $\{ k_1, \ldots , k_j \}$ containing $j$ elements of the set $\mathcal{S}^{(N)}$. We recall that all elements of a set are by construction distinct (see e.g. \cite{Rys}), so we must have $k_{j'} \neq k_{j''}$. This is precisely ensured by the fact that the summation indices $k_1, \ldots , k_j$ in~\eqref{P_j_N_prod_state} are required to satisfy $k_2>k_1, \ldots , k_j>k_{j-1}$. Since the total number of subsets containing $j$ elements of a set of $N$ elements is known \cite{Rys} to merely be the binomial coefficient $\binom{N}{j} \equiv N!/j!(N-j)!$, we have
\begin{align}
\sum_{k_1=1}^{N} \sum_{\substack{k_2=1 \\ k_2>k_1}}^{N} \cdots \sum_{\substack{k_j=1 \\ k_j>k_{j-1}}}^{N} 1 = \binom{N}{j} \, .
\label{nested_sum}
\end{align}

Therefore, substituting~\eqref{nested_sum} into~\eqref{P_j_N_prod_state} and recognizing the definitions~\eqref{P_1_1_def} and~\eqref{P_0_1_def} of the single-particle probabilities $\mathcal{P}_{1}^{(1)}$ and $\mathcal{P}_{0}^{(1)}$, respectively, shows that $\mathcal{P}_{j}^{(N)}$ can be written in the form
\begin{align}
\mathcal{P}_{j}^{(N)}(t) = \binom{N}{j} \left[ \mathcal{P}_{1}^{(1)}(t) \right]^{j} \left[ \mathcal{P}_{0}^{(1)}(t) \right]^{N-j} \, ,
\label{P_j_N_factorized}
\end{align}
for any $j=0, \ldots , N$. Setting in particular $j=0$ into~\eqref{P_j_N_factorized} yields
\begin{align}
\mathcal{P}_{0}^{(N)}(t) = \left[ \mathcal{P}_{0}^{(1)}(t) \right]^{N} \, ,
\label{P_0_N_factorized}
\end{align}
so that we get for the probability $\mathcal{P}_{-}^{(N)}$, after substituting~\eqref{P_0_N_factorized} into~\eqref{P_BF_P_0_N_rel},
\begin{align}
\mathcal{P}_{-}^{(N)}(t) = 1 - \left[ \mathcal{P}_{0}^{(1)}(t) \right]^{N} \, .
\label{P_BF_P_0_1_rel}
\end{align}
Furthermore, substituting~\eqref{P_0_N_factorized} into~\eqref{Delta_N_def_P_0} yields for the probability change $\Delta_N$
\begin{align}
\Delta_N = \left[ \mathcal{P}_{0}^{(1)}(0) \right]^N - \left[ \mathcal{P}_{0}^{(1)}(T) \right]^N \, .
\label{Delta_N_P_0_factorized}
\end{align}

It is worth noting that the form~\eqref{P_BF_P_0_1_rel} ensures that the general structure of $\mathcal{P}_{-}^{(N)}$ is the same as that of $\mathcal{P}_{-}^{(1)}$, for any $N \geqslant 2$. Indeed, differentiating~\eqref{P_BF_P_0_1_rel} with respect to the time $t$ yields
\begin{align*}
\frac{d \mathcal{P}_{-}^{(N)}}{dt} = - N \left[ \mathcal{P}_{0}^{(1)}(t) \right]^{N-1} \frac{d \mathcal{P}_{0}^{(1)}}{dt} \, ,
\end{align*}
that is, since $d \mathcal{P}_{0}^{(1)} / dt = - d \mathcal{P}_{1}^{(1)} / dt$ and $\mathcal{P}_{-}^{(1)} = \mathcal{P}_{1}^{(1)}$ by construction,
\begin{align}
\frac{d \mathcal{P}_{-}^{(N)}}{dt} = N \left[ \mathcal{P}_{0}^{(1)}(t) \right]^{N-1} \frac{d \mathcal{P}_{-}^{(1)}}{dt} \, .
\label{dP_BF_dt_expr}
\end{align}
Now, the probability $\mathcal{P}_{0}^{(1)}(t)$ is positive and generally non zero at finite times $t \geqslant 0$. Actually, if $\mathcal{P}_{0}^{(1)}(t)=0$ at some time $t$, then no backflow can occur at immediate subsequent times since in such a case the probability $\mathcal{P}_{0}^{(1)}$ can not decrease. We can thus readily see on~\eqref{dP_BF_dt_expr} that the maxima and minima of $\mathcal{P}_{-}^{(N)}$ precisely correspond to those of $\mathcal{P}_{-}^{(1)}$, for any $N \geqslant 2$.

The factorized form~\eqref{Delta_N_P_0_factorized} of $\Delta_N$, which we emphasize stems from the expression~\eqref{psi_N_prod_state_def} of the $N$-particle wave function $\psi^{(N)}$ as a product state, can then be adequately used to study the dependence of the phenomenon of quantum backflow with respect to the number $N$ of bosons, as we now discuss.


\subsection{Quantum backflow in the limit \texorpdfstring{$N \to \infty$}{N infinite}}
\label{class_limit_subsec}

Being a probability, $\mathcal{P}_{0}^{(1)}(t)$ takes values between 0 and 1 at any time $t$. Since the expression~\eqref{Delta_N_P_0_factorized} of $\Delta_N$ involves the difference of $N^{\text{-th}}$ powers of $\mathcal{P}_{0}^{(1)}$, it should be clear that $\Delta_N$ becomes arbitrarily small as $N$ increases if $\mathcal{P}_{0}^{(1)}(0),\mathcal{P}_{0}^{(1)}(T) \neq 1$. In order to make this precise and derive quantitative bounds for $\Delta_N$, we first factorize~\eqref{Delta_N_P_0_factorized} by the single-particle probability change $\Delta_1 = \mathcal{P}_{0}^{(1)}(0) - \mathcal{P}_{0}^{(1)}(T)$ and we have
\begin{align}
\Delta_N = \left\{ \sum_{k=0}^{N-1} \left[ \mathcal{P}_{0}^{(1)}(0) \right]^k \left[ \mathcal{P}_{0}^{(1)}(T) \right]^{N-1-k} \right\} \Delta_1 \, .
\label{Delta_N_Delta_1_rel}
\end{align}
Note that to have $\mathcal{P}_{0}^{(1)}(0)=\mathcal{P}_{0}^{(1)}(T)=0$ readily yields $\Delta_N=\Delta_1=0$ in view of~\eqref{Delta_N_P_0_factorized}. Therefore, it is clear on~\eqref{Delta_N_Delta_1_rel} that $\Delta_N$ is strictly positive if and only if $\Delta_1$ is. In other words, the $N$-particle product state $\psi^{(N)}$ given by~\eqref{psi_N_prod_state_def} gives rise to the phenomenon of quantum backflow if and only if the single-particle state $\psi^{(1)}$ does.

Now, suppose that backflow occurs for the single-particle state $\psi^{(1)}$, i.e. we have $\Delta_1 > 0$ and thus in view of~\eqref{Delta_N_P_0_factorized} for $N=1$
\begin{align}
\mathcal{P}_{0}^{(1)}(T) < \mathcal{P}_{0}^{(1)}(0) \, .
\label{BF_cond_P_0}
\end{align}
We hence have as a direct consequence of~\eqref{BF_cond_P_0} that
\begin{align}
\sum_{k=0}^{N-1} \left[ \mathcal{P}_{0}^{(1)}(0) \right]^k \left[ \mathcal{P}_{0}^{(1)}(T) \right]^{N-1-k} < N \left[ \mathcal{P}_{0}^{(1)}(0) \right]^{N-1} \, .
\label{ineq_prod_P_0}
\end{align}
Combining~\eqref{Delta_N_Delta_1_rel} with~\eqref{BF_cond_P_0}-\eqref{ineq_prod_P_0} hence yields the following inequality satisfied by $\Delta_N$:
\begin{align}
0 < \Delta_N < a_N \, \Delta_1 \, ,
\label{Delta_N_Delta_1_up_bound}
\end{align}
where we introduced the quantity $a_N$ defined by
\begin{align}
a_N \equiv N \left[ \mathcal{P}_{0}^{(1)}(0) \right]^{N-1} \, .
\label{a_N_def}
\end{align}
Note that $a_N$ only depends on $N$ and on the initial probability $\mathcal{P}_{0}^{(1)}(0)$, and is thus in particular independent of the duration $T$.

We now assume that
\begin{align}
\mathcal{P}_{0}^{(1)}(0) < 1 \, ,
\label{P_0_less_1}
\end{align}
even though $\mathcal{P}_{0}^{(1)}(0)$ can be arbitrarily close to 1. We then rewrite the quantity~\eqref{a_N_def} in the form
\begin{align}
a_N = \frac{1}{\mathcal{P}_{0}^{(1)}(0)} N \, \mathrm{exp} \left\{- \left\lvert \mathrm{ln} \left[ \mathcal{P}_{0}^{(1)}(0) \right] \right\rvert N \right\} \, .
\label{a_N_exp_form}
\end{align}
We emphasize that to divide by or to take the logarithm of $\mathcal{P}_{0}^{(1)}(0)$ is ensured by the fact that $\mathcal{P}_{0}^{(1)}(0) \neq 0$. Indeed, to have $\mathcal{P}_{0}^{(1)}(0)=0$ would contradict the hypothesis~\eqref{BF_cond_P_0} of the presence of backflow, as it would then yield a negative probability $\mathcal{P}_{0}^{(1)}(T)$. Now, the assumption~\eqref{P_0_less_1} ensures that the logarithm in~\eqref{a_N_exp_form} does not vanish. We hence get in the limit $N \to \infty$
\begin{align}
\lim_{N \to \infty} a_N = 0 \, .
\label{limit_a_N}
\end{align}
Finally, taking the limit $N \to \infty$ in~\eqref{Delta_N_Delta_1_up_bound} readily yields, in view of~\eqref{limit_a_N} and using the squeeze theorem,
\begin{align}
\lim_{N \to \infty} \Delta_N = 0 \, ,
\label{limit_Delta_N}
\end{align}
as anticipated.

Our analysis hence shows that, in the case of $N$ bosons in a same one-particle state, increasing the number $N$ of bosons makes the maximum achievable backflow $\Delta_{N,\text{max}}$ become arbitrarily small. That is, we analytically showed that the phenomenon of quantum backflow vanishes in the limit $N \to \infty$ for this class of many-particle systems. We believe that this provides an alternative insight regarding the classical limit of the fundamentally quantum phenomenon of backflow, whose magnitude is thus seen to decrease when the system reaches a sufficiently large size. This strongly suggests that to observe this phenomenon on a macroscopic system is basically impossible.

To conclude this subsection, we briefly discuss the accuracy of the inequality~\eqref{Delta_N_Delta_1_up_bound}. We first note that the lower bound in~\eqref{Delta_N_Delta_1_up_bound} can be easily refined. Indeed, as is recalled in section~\ref{single_part_sec} above, the single-particle probability change $\Delta_1$ is bounded by the Bracken-Melloy constant~\eqref{BM_constant}. We hence have $\Delta_{1} \leqslant \Delta_{1,\text{max}}$, that is in view of~\eqref{Delta_N_P_0_factorized} for $N=1$
\begin{align}
\mathcal{P}_{0}^{(1)}(T) \geqslant \mathcal{P}_{0}^{(1)}(0) - \Delta_{1,\text{max}} \, ,
\label{lower_bound_P_0_T}
\end{align}
and thus
\begin{align}
\sum_{k=0}^{N-1} \left[ \mathcal{P}_{0}^{(1)}(0) \right]^k \left[ \mathcal{P}_{0}^{(1)}(T) \right]^{N-1-k} \geqslant b_N \, ,
\label{lower_bound_sum}
\end{align}
where the quantity $b_N$ is defined by
\begin{align}
b_N \equiv \sum_{k=0}^{N-1} \left[ \mathcal{P}_{0}^{(1)}(0) \right]^k \left[ \mathcal{P}_{0}^{(1)}(0) - \Delta_{1,\text{max}} \right]^{N-1-k} \, .
\label{b_N_def}
\end{align}
Similarly to $a_N$, $b_N$ only depends on $N$ and $\mathcal{P}_{0}^{(1)}(0)$, and is independent of $T$. Combining~\eqref{Delta_N_Delta_1_rel} with~\eqref{lower_bound_sum} then shows, also using~\eqref{Delta_N_Delta_1_up_bound}, that $\Delta_N$ satisfies the inequality
\begin{align}
b_N \, \Delta_1 \leqslant \Delta_N < a_N \, \Delta_1 \, .
\label{Delta_N_inequality}
\end{align}

Note that in view of its definition~\eqref{b_N_def} the quantity $b_N$ is by construction an expansion in powers of $\Delta_{1,\text{max}}$, and we have with~\eqref{a_N_def}
\begin{align}
b_N = a_N - \frac{N(N-1)}{2} \left[ \mathcal{P}_{0}^{(1)}(0) \right]^{N-2} \Delta_{1,\text{max}} + \ldots \, .
\end{align}
Since the Bracken-Melloy constant takes the relatively small numerical value $\Delta_{1,\text{max}} \approx 0.0384517$ [see~\eqref{BM_constant}], we hence generally have $b_N \approx a_N$. The bounds in the inequality~\eqref{Delta_N_inequality} are thus expected to be rather tight in general. In particular, only in those cases where $\mathcal{P}_{0}^{(1)}(0) < \Delta_{1,\text{max}}$ may the quantity $b_N$ defined by~\eqref{b_N_def} be negative, hence making the original inequality~\eqref{Delta_N_Delta_1_up_bound} possibly stronger than the refined one~\eqref{Delta_N_inequality}.

Our conclusions are valid for an arbitrary product state~\eqref{psi_N_prod_state_def}, as long as the condition~\eqref{P_0_less_1} is satisfied. We now illustrate them on an explicit example.


\subsection{Explicit example}
\label{example_N_bos_subsec}

We conclude this paper by illustrating some of the above-discussed features of $N$-boson quantum backflow on the explicit example outlined in subsection~\ref{example_subsec}. That is, we assume that the single-particle state is $\psi^{(1)}(x,t) = \widetilde{\psi}(x',t')$ with $\widetilde{\psi}$ given by~\eqref{psi_t_error_fct_expr} and the dimensionless variables $x'$ and $t'$ defined by~\eqref{dimensionless_x_and_t_def}. The $N$-boson wave function~\eqref{psi_N_prod_state_def} in this case hence reads
\begin{align}
\psi^{(N)}(\bol{x},t) = \prod_{j=1}^{N} \widetilde{\psi}\left( \frac{\alpha x_j}{\hbar}, \frac{\alpha^2 t}{m \hbar} \right) \, ,
\label{psi_N_example}
\end{align}
while the one-particle probabilities $\mathcal{P}_{0,1}^{(1)}(t)=\widetilde{\mathcal{P}}_{0,1}^{(1)}(t')$ are given by~\eqref{P_1_1_example}-\eqref{P_0_1_example}. It is worth noting that the initial one-particle probabilities can be shown to be equal and we hence have here
\begin{align}
\widetilde{\mathcal{P}}_{1}^{(1)}(0) = \widetilde{\mathcal{P}}_{0}^{(1)}(0) = \frac{1}{2} \, .
\label{init_one_part_proba_expr}
\end{align}
Substituting~\eqref{P_1_1_example}-\eqref{P_0_1_example} into~\eqref{P_0_N_factorized} and~\eqref{P_BF_P_0_1_rel} then readily yields the corresponding expressions of the $N$-particle probabilities $\widetilde{\mathcal{P}}_{0}^{(N)}(t')$ and $\widetilde{\mathcal{P}}_{-}^{(N)}(t')$, respectively.


\begin{figure}[ht]
\centering
\includegraphics[width=0.48\textwidth]{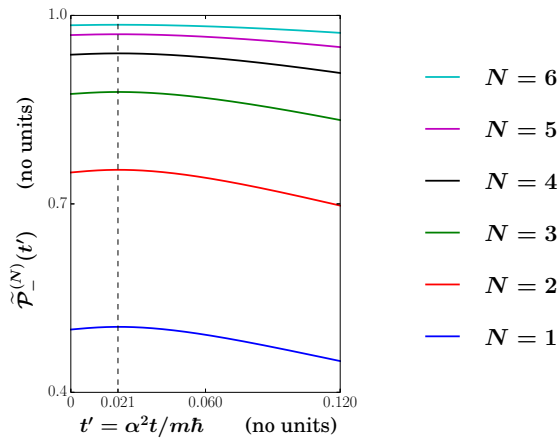}
\caption{(Color online) Probability $\widetilde{\mathcal{P}}_{-}^{(N)}$ as a function of the dimensionless time parameter $t'$ [defined by~\eqref{dimensionless_x_and_t_def}] for different numbers $N$ of bosons, from $N=1$ (blue) to $N=6$ (cyan).}
\label{P_-_fig}
\end{figure}


Figure~\ref{P_-_fig} shows the probability $\widetilde{\mathcal{P}}_{-}^{(N)}$ as a function of the dimensionless time parameter $t'$ [related to the physical time $t$ through~\eqref{dimensionless_x_and_t_def}] for different numbers $N$ of bosons. The blue curve corresponds to $N=1$ and illustrates the known fact \cite{BM94}, recalled in subsection~\ref{example_subsec} above, that the derivative $d \widetilde{\mathcal{P}}_{1}^{(1)}/dt'$ remains positive for $t'$ ranging between 0 and $t'_1 \approx 0.021$. Indeed, we can readily check that $\widetilde{\mathcal{P}}_{-}^{(1)}$, i.e. merely $\widetilde{\mathcal{P}}_{1}^{(1)}$ in view of~\eqref{P_min_def}, reaches a maximum at $t'=t'_1$. The dashed black vertical line located at $t'=0.021$ then highlights the fact that the probabilities $\widetilde{\mathcal{P}}_{-}^{(N)}$ for $N \geqslant 2$ are also maximum at the same time $t'=t'_1$ (as we explicitly checked on the data). This is an illustration of the particular structure of $\widetilde{\mathcal{P}}_{-}^{(N)}$ as a function of $t'$ that is embedded in~\eqref{dP_BF_dt_expr}.

As is clear on figure~\ref{P_-_fig}, the probability $\widetilde{\mathcal{P}}_{-}^{(N)}(t')$ increases with $N$ at any fixed time $t'$. This is expected as it is by construction the probability of finding at least one boson on the negative real axis. To increase the number $N$ of bosons hence also increases the number of events that contribute to this probability. However, we emphasize that this increase of $\widetilde{\mathcal{P}}_{-}^{(N)}$ with $N$ does \textit{not} mean that quantum backflow itself increases with $N$ as well. Indeed, the latter is characterized by the increase of $\widetilde{\mathcal{P}}_{-}^{(N)}$ over a certain time interval at fixed $N$.

\begin{figure}[ht]
\centering
\includegraphics[width=0.48\textwidth]{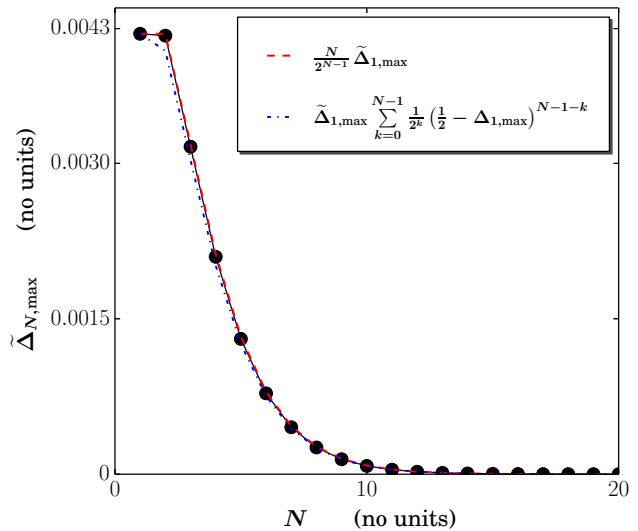}
\caption{(Color online) Maximum increase $\widetilde{\Delta}_{N,\text{max}}$, given by~\eqref{Delta_N_max_example_def}, of the probability $\widetilde{\mathcal{P}}_{-}^{(N)}$ as a function of the number $N$ of bosons (black disks joined by the solid black line), along with the lower (dash-dotted blue curve) and upper (dashed red curve) bounds of the inequality~\eqref{Delta_N_max_ineq}.}
\label{Delta_N_max_fig}
\end{figure}

In the present case the probabilities $\widetilde{\mathcal{P}}_{-}^{(N)}$ are increasing functions from $t'=0$ to $t'=t'_1$, and decreasing functions for $t'>t'_1$. Therefore, the corresponding maximum backflow $\widetilde{\Delta}_{N,\text{max}}$ is here merely given by
\begin{align}
\widetilde{\Delta}_{N,\text{max}} = \widetilde{\mathcal{P}}_{-}^{(N)}(t'_1) - \widetilde{\mathcal{P}}_{-}^{(N)}(0) \, .
\label{Delta_N_max_example_def}
\end{align}
The behavior of the latter quantity with respect to the number $N$ of bosons is illustrated on figure~\ref{Delta_N_max_fig} by the black disks (joined by the solid black line). First, we find for $N=1$ a value $\widetilde{\Delta}_{1,\text{max}} \approx 0.00425$, in agreement with the value~\eqref{Delta_expr} originally obtained in \cite{BM94}. We can then readily see that $\widetilde{\Delta}_{N,\text{max}}$ reaches a value as small as $\widetilde{\Delta}_{20,\text{max}} \approx 1.5 \times 10^{-7}$, i.e. $\widetilde{\Delta}_{20,\text{max}} \approx 3.53 \times 10^{-5} \, \widetilde{\Delta}_{1,\text{max}}$, for $N=20$.


Furthermore, combining the general inequality~\eqref{Delta_N_inequality} with the expression~\eqref{init_one_part_proba_expr} of the initial probability $\widetilde{\mathcal{P}}_{0}^{(1)}(0)$ and the definitions~\eqref{a_N_def} and~\eqref{b_N_def} of $a_N$ and $b_N$, respectively, shows that $\widetilde{\Delta}_{N,\text{max}}$ satisfies
\begin{multline}
\widetilde{\Delta}_{1,\text{max}} \sum\limits_{k=0}^{N-1} \frac{1}{2^k} \left( \frac{1}{2} - \Delta_{1,\text{max}} \right)^{N-1-k} \\
\leqslant \widetilde{\Delta}_{N,\text{max}} < \frac{N}{2^{N-1}} \, \widetilde{\Delta}_{1,\text{max}} \, .
\label{Delta_N_max_ineq}
\end{multline}
We recall that the (approximate) value of the Bracken-Melloy constant $\Delta_{1,\text{max}}$ is given by~\eqref{BM_constant}. The lower and upper bounds of~\eqref{Delta_N_max_ineq} are represented by the dash-dotted blue and dashed red, respectively, curves on figure~\ref{Delta_N_max_fig}. We can thus readily see that the inequality~\eqref{Delta_N_max_ineq} is indeed rather tight.


\section{Conclusion}
\label{A}

In this paper we investigated how the phenomenon of quantum backflow extends to many-particle systems. We considered a system formed of $N \geqslant 1$ identical nonrelativistic structureless free particles. Our formulation of many-particle quantum backflow is then based on the change $\Delta_N$ [defined by~\eqref{Delta_N_def} or equivalently~\eqref{Delta_N_def_P_0}] of the probability $\mathcal{P}_{-}^{(N)}$ [defined by~\eqref{P_min_def}] of finding \textit{at least} one particle on the negative real axis over a fixed but arbitrary time interval $0 \leqslant t \leqslant T$, for some $T>0$. Similarly to the single-particle case, backflow occurs whenever $\Delta_N > 0$.

We then saw how our general formulation of many-particle quantum backflow, valid for either bosons or fermions as well as for distinguishable particles, greatly simplifies in the particular case of a system composed of $N$ identical bosons that are all in the same one-particle state. The $N$-particle wave function $\psi^{(N)}$ can thus be written as the mere product state~\eqref{psi_N_prod_state_def}. We showed in this case that the maximum achievable backflow $\Delta_{N,\text{max}}$ becomes arbitrarily small as the number $N$ of bosons increases, which is the outcome of Eq.~\eqref{limit_Delta_N}. We emphasize that this result is exact and did not require any numerical analysis. This alternative approach to the classical limit of quantum backflow hence seems to confirm our physical intuition that this intrinsically quantum phenomenon vanishes for a large system.

Many-particle quantum backflow spans a vastly uncharted territory, as the current understanding of this effect has been entirely built on single-particle systems. We hence believe that our study opens up various prospects for further research. For instance, while we showed with~\eqref{limit_Delta_N} that $\Delta_N$ vanishes, for a $N$-boson system in the product state~\eqref{psi_N_prod_state_def}, in the limit $N \to \infty$, nothing \textit{a priori} precludes the fact that the maximum backflow $\Delta_{N,\text{max}}$ may actually increase over some finite range of values of $N$. It would thus be interesting to investigate whether or not this is the case by explicitly computing $\Delta_{N,\text{max}}$ e.g. for the lowest values of $N$. This could in particular allow to refine the general inequality~\eqref{Delta_N_inequality} that we obtained here by providing a better estimation of the corresponding least upper bound. An other prospect for deepening our understanding of many-particle quantum backflow would be to consider more general $N$-particle wave functions than the product state~\eqref{psi_N_prod_state_def}. Indeed, our assumption of a system formed of $N$ bosons in a same one-particle state, though practically relevant, restricts the space of positive-momentum states of the form~\eqref{psi_N_t_gen_expr} that we consider. The precise impact of the nature, bosonic or fermionic, of the particles on quantum backflow is yet an other potentially promising avenue. We hope that our work can pave the way towards a closer investigation of these, among others, aspects.


\section*{Acknowledgments}

I am grateful to P. Gaspard and A. Goussev for valuable comments. I also thank J. Hurst for motivating discussions. I acknowledge financial support by the Universit\'e Libre de Bruxelles (ULB) and the Fonds de la Recherche Scientifique~-~FNRS under the Grant PDR~T.0094.16 for the project ``SYMSTATPHYS".





\bibliography{DATABASE_Quantum_Backflow}


\end{document}